\documentclass[10pt,a4paper]{article} 
\begin{document}

\title{Damping of Gravitational Waves and Density Perturbations in the Early Universe}
\author{A. Dimitropoulos\thanks{e-mail address: A.Dimitropoulos@astro.cf.ac.uk}\\University 
of Wales, College of Cardiff,\\ Department of Physics and Astronomy,\\  P.O.Box 913, Cardiff, CF2 3YB, United Kingdom}
\maketitle

\begin{abstract}
Since the discovery of the large angular scale anisotropies in the microwave background radiation, the behaviour of cosmological perturbations (especially, density 
perturbations and gravitational waves) has been of great interest. In this study, after a detailed and rigorous treatment of the behaviour of 
gravitational waves in viscous cosmic media, we conclude that the damping of cosmological gravitational waves of long wavelengths is negligible for most cases of physical 
interest. A preliminary analysis suggests that similar results hold for density perturbations in the long wavelength limit. Therefore, long wavelength cosmological perturbations have not been practically 
affected by viscous processes, and are good probes of the very early Universe.
\end{abstract}

\section{Introduction}

A likely explanation of the observed large angular scale anisotropy of the Cosmologigal Microwave Background Radiation (CMBR) is 
cosmological perturbations of quantum mechanical origin (Grishchuk 1993), mainly gravitational waves (Grishchuk 1994). However, one could argue that these 
perturbations are not responsible for the CMBR anisotropy if they could be washed out by viscous processes. As a first step, one should 
investigate whether, in a classical approach, the amplitude of such perturbations decreases significantly or not. This is the point of this study.We 
have dealt mainly with gravitational waves in several viscous cosmic media, possesing only shear viscosity. One could refer to some standard sources,like 
Hawking 1966, Weinberg 1972 and Grishchuk and Polnarev 1980. We adopted a technique previously used by 
Weinberg 1972. Several particular models for viscous cosmic media were taken from Mendez et al 1997.

The assumptions used in the study of the early Universe can be found in standard text books, for example Weinberg 1972. Our study indicated that the zero chemical 
potential approximation should be reconsidered by the end of the radiation dominated era, in order to get correct decoupling time between matter and 
radiation, and the hydrogen and baryon abundances should be taken into account. We neglect the bulk viscosity.

Information concerning transport phenomena can be found in Landau and Lifshitz 1966, Tabor 1970 and Reif 1965. We will only include the information necessary 
for our study.
By "viscosity" we will mean specifically shear viscosity. If $\tau_{i}$ is the mean free time between 
collisions, $\epsilon_{t.i}$ is the thermal energy density of the particles of the i species responsible for momentum transfers, then the shear viscosity 
coefficient is 
   \begin{equation}
     \xi=\theta \sum_{i}\epsilon_{t.i}\tau_{i}   \label{eq:eva}
   \end{equation}
where $\theta$ is a numerical constant.
 
Elementary information required about Local Thermodynamical Equilibrium, Dissipation and Decoupling are presented in Table 1 (Weinberg 1972, 
Kolb and Turner 1990). We should mention though that these conditions are a consequence of the constancy of the entropy per comoving volume element.  
We define a characteristic Hubble time by \( T_{H}\equiv \mathcal{R}/\dot{\mathcal{R}} \) 
where $\mathcal{R}$ is the scale factor.
$\eta$ is conformal time, a dot is differentiation with respect to t, while a prime with respect to $\eta$.

   \begin{table}
      \begin{tabular}{|c|c|}   \hline
      $\tau \ll T_{H}$   &Perfect thermal equilibrium holds.Dissipation is negligible.  \\  \hline
      $\tau \leq T_{H}$  &Departures from thermal equilibrium begin.At the equality sign, decoupling occurs and \\
                         &this interaction is no longer realised.Dissipation becomes important,untill of course \\
                         &decoupling occurs and the interaction ceases.  \\   \hline
      $\tau > T_{H}$     &Decoupling has occured, and no dissipation takes place since the interaction has stopped. \\   \hline
      \end{tabular}
      \caption{LOCAL THERMAL EQUILIBRIUM, DISSIPATION AND DECOUPLING}
   \end{table}
   
\section{The behaviour of Gravitational Waves in a non viscous medium}
\label{sec-behano}

Lifshitz pointed out  how the different types of perturbations can be constructed in the form of scalar, vector and 
tensor harmonics, corresponding to density perturbations, rotational perturbations and gravitational waves respectively (see,for example, Lifshitz and 
Khalatnikov 1963). The perturbed Einstein equations are given in Weinberg 1972. For gravitational waves, $h_{i}^{j}$ denotes the time dependent part of the perturbations and q the constant wavenumber of the perturbation, related to the time 
dependent wavelength $\lambda(t)$ by the relation \( q \equiv \frac{2\pi\mathcal{R}}{\lambda} \). Since all the $h_{i}^{j}$ components obey the 
same equations, we will ignore the indices in our notation and refer to a single component, as h. This equation, has been transformed and interpreted as a parametrically excited oscillator, 
(Grishchuk 1993). 
The $\mu(\eta)$ amplitude is related to the $h(\eta)$ amplitude by  
\( h(\eta)=\mu(\eta)\mathcal{R}(\eta)^{-1} \). The $\mathcal{R}^{-1}$ variation 
of h reflects the adiabatic decrease of h. The perturbations interact with the background time dependent gravitational field, which supplies energy to waves with 
wavelengths that satisfy the parametric amplification condition. In the case of gravitational waves, the interaction potential \( U(\eta) \equiv \frac{\mathcal{R}''(\eta)}{\mathcal{R}(\eta)} \) 
represents the background gravitational 
field. In order for this interaction to take place, the frequency of the wave must be comparable with that of the variations of the background field. 
Depending on the wavelengths of the perturbations, the behaviour of $\mu$ and h is as described in Table 2.
Therefore, gravitational waves interact parametricaly with the background gravitational field, and the ones longer than the Hubble radius are "superadiabatically"amplified.
The quantum treatment of this 
phenomenon and its implications to the CMBR statistics and anisotropy is investigated in Grishchuk 1993.

\begin{table}
      \begin{tabular}{|c|c|}   \hline
      $q \gg U(\eta)$  &In this overbarrier region, the oscillatory solutions \\
                       &experience only adiabatic decrease, since no \\ 
                       &interaction with the barrier takes place.The \\
                       &expressions for the amplitudes are  \\
                       &\( \mu(\eta)=C_{1}\exp \{\pm iq\eta\} \) and \\
                       &\( h(\eta)=C_{1}\mathcal{R}^{-1}\exp \{\pm iq\eta\} \). \\  \hline
      $q \ll U(\eta)$  &In this underbarrier region, the solutions are practically \\
                       &constant.These long wavelengths do not suffer from adiabatic \\
                       &decrease: due to their interaction with the barrier they are \\ 
                       &amplified ("superadiabatic amplification").The expressions \\
                       &for the amplitudes are \\
                       &\( \mu(\eta)=C_{b}\mathcal{R}+C_{a}\mathcal{R}\int \mathcal{R}^{-2}d(\eta) \simeq C_{b}\mathcal{R} \) \\
                       &and \( h(\eta)=C_{b}+C_{a}\int \mathcal{R}^{-2}d(\eta) \simeq C_{b} \).\\   \hline
      \end{tabular}
      \caption{AMPLITUDE OF GRAVITATIONAL WAVES IN THE NONVISCOUS CASE}
\end{table}

\section{The behaviour of Gravitational Waves in the presence of a Shear Viscosity}

In order to put the equations of propagation for gravitational waves in a viscous medium in a form that reveals the underlying physics, we will first derive 
an expression for shear viscosity. We begin from equation (\ref{eq:eva}). It is obvious that species with $\epsilon_{t.i}\tau_{i}$ several orders of magnitudes 
smaller than those of the rest constituents of the fluid will not participate. We will make some assumptions. The first is that in the radiation dominated 
era, the thermal energy density of a species is approximately equal to its total energy density: $\epsilon_{t.i} \simeq \epsilon_{i}$. The second assumption 
is that the $\epsilon_{i}$ of the different species are of the same order of magnitude. We define the number $\kappa_{i} \equiv \epsilon_{i} / \epsilon$, where 
epsilon is the total energy density of the fluid. The different $\kappa_{i}$ will be of the same order of magnitude, and will all be denoted by $\kappa$. 
Therefore, (\ref{eq:eva}) reduces to
    \begin{equation}
        \xi = \theta (\sum_{i} \kappa_{i} \tau_{i}) \epsilon     \label{eq:evatovos}
    \end{equation}
Obviously, the quantity $\bar{\tau} \equiv \sum_{i} \kappa_{i} \tau_{i})$ is the mean free time of the fluid, and will be at most of order of the largest 
$\tau_{i}$. The third assumption, is that the $\tau_{i}$ of the species contributing to $\xi$ are of the same order of magnitude and equal to $\tau$. Then, 
(\ref{eq:evatovos}) becomes
     \begin{equation}
         \xi = \theta \psi \kappa \epsilon \tau     \label{eq:evadis}
     \end{equation}
The fourth assumption is that the number of species contributing to viscosity is small: $\psi \sim 1$.

Following Weinberg 1972, one finds that in the presence of a cosmic medium possesing shear viscosity the propagation equations for the gravitational waves' 
amplitude is 
    \begin{equation}
       \ddot{h}+(3T_{H}^{-1}+6\theta \psi \kappa T_{H}^{-2}\tau)\dot{h}+(\frac{cq}{\mathcal{R}})^{2}h=0    \label{eq:tessera}
    \end{equation}
where we have used (\ref{eq:evadis}) and expressed $\epsilon$ in terms of $T_{H}$ from the unperturbed Einstein equations. Equation~(\ref{eq:tessera}) 
shows that a further (above adiabatic damping) decrease of the amplitude arises due to shear viscosity. This equation in terms of conformal time and for 
the $\mu$ amplitude is 
    \begin{equation}
       \mu''+6\theta \psi \kappa A^{2}\frac{c\tau}{\mathcal{R}}\mu'+(q^{2}-V(\eta))\mu=0    \label{eq:avwvumn}
    \end{equation}
where \( A\equiv\frac{\mathcal{R}'}{\mathcal{R}} \) and the time dependent potential is \( V(\eta)=U(\eta)+6\theta \psi \kappa A^{3}\frac{c\tau}{\mathcal{R}} \). 
We put (\ref{eq:avwvumn}) into the form of the Schroedinger equation by introducing the function m:  
    \begin{equation}
       \mu=m\exp \{-\int 3\theta \psi \kappa A^{2}\frac{c\tau}{\mathcal{R}}d(\eta)\}     \label{eq:karavwvumn}    
    \end{equation}
Then, equation~(\ref{eq:karavwvumn}) becomes
    \begin{equation}
       m''+(q^{2}-Y(\eta))m=0    \label{eq:pevte}
    \end{equation}
with $Y(\eta)=F'+F^{2}$ where \( F=A+3\theta \psi \kappa A^{2}\frac{c\tau}{\mathcal{R}} \). This is again the equation for a parametrically excited 
oscillator, but with a modified potential due to shear viscosity.

It is known that the dissipation is expected to be negligible when the mean free time between collisions is much less than the Hubble
time. This is a consequence of the constancy of entropy within a comoving volume element, which yields the relation $\mathcal{R} \propto T^{-1}$. The expansion
rate $T_{H}^{-1}$ determines the rate of temperature change: particles that are in thermal equilibrium should have an interaction rate greater than the rate of 
temperature change. For $\tau \propto T^{-q}$, like the ones we will consider, when $\tau \geq T_{H}$, a particle will interact less than once, so this species 
will drop out of equilibrium.

When $\tau < T_{H}$, the viscous term participating in the cofactor of $\dot{h}$ in (\ref{eq:tessera}) is much less than the expansion term. Even 
in the case of non negligible viscosity, the viscous term can never become larger than the expansion term, because otherwise the condition of thermal equilibrium 
(not necessarily perfect thermal equilibrium) (see Table~1 ) will be violated. This means that decoupling will occur 
and the dissipative mechanism under consideration will cease to function as such. Significant dissipation is expected to occur at those times when the viscous 
term becomes comparable to the expansion term: then, we have large departures from perfect thermal equilibrium.

Let $\hat{h}$ be a solution in the absence of viscosity and $\tilde{h}$ a solution of (\ref{eq:tessera}), that is, in the presence of viscosity. A measure of 
the dissipation can be presented as 
    \begin{equation}
       Z=\frac{\hat{h}-\tilde{h}}{\hat{h}}  \label{eq:eksi}
    \end{equation}

We will now consider short and long wavelengths separately.

\subsection{Solution for short wavelengths}

A solution of (\ref{eq:tessera}) in this limiting case of short waves is (Weinberg 1972)
    \begin{equation}
        \tilde{h}=\hat{h}\exp \{-\int 3\theta \psi \kappa T_{H}^{-2}\tau dt\}   \label{eq:epta}
    \end{equation}
This solution is applicable under the condition \( T_{H} \ll \frac{cq}{\mathcal{R}} \) that is, when the relevant waves are well inside the (time dependent) 
Hubble radius. This result can be obtained from (\ref{eq:pevte}) by neglecting the 
potential $Y(\eta)$. Combining (\ref{eq:eksi}) with (\ref{eq:epta}), one derives for this limiting case the damping 
    \begin{equation}
       Z=1-\exp \{-\int 3\theta \psi \kappa T_{H}^{-2}\tau dt\} \label{eq:oktw}
    \end{equation}
There is no a priori reason for this damping to be much smaller than 1, nevertheless, it will be at most of order unity, otherwise, the condition of thermal 
equilibrium will be violated.

\subsection{Solution for long wavelengths}

This is the limit which, apparently, was not considered before for viscous matter. In this case, wavelengths are much longer than the Hubble radius in the 
era under consideration and, therefore, the $q^{2}$ term in (\ref{eq:pevte}) can be neglected.
We have found a solution of equations (\ref{eq:pevte}) and (\ref{eq:tessera}) in the form
    \begin{equation}
        \tilde{h}=C_{1}+C_{2}\int\mathcal{R}^{-3}(\exp \{-2\int 3\theta \psi \kappa T_{H}^{-2}\tau dt\})dt   \label{eq:evvea}
    \end{equation}
The solution in the nonviscous case is \( \hat{h}=C_{1}+C_{2}\int \mathcal{R}^{-3}dt \) so viscosity affects only the 
"decaying" second term, but not the constant ("growing") first term. Even in the absence of viscosity, the "decaying" term is much smaller than the "growing"
 one. 
Thus, since the affected by viscosity "decaying" term in $\tilde{h}$ is smaller, or, at most, equal to the "decaying" term in $\hat{h}$, it will also be much 
smaller than the (unaffected) constant term. Therefore, 
\[ \tilde{h} \approx C_{1} \approx \hat{h} \] This means that the viscosity does not practically affect the amplitude of waves with wavelengths longer
than the Hubble radius. Of course, the same holds for the wavelengths that are longer than the Hubble radius 
today.

This is a central result of our study, because these long wavelengths are responsible for the large angular 
scale CMBR anisotropy. In this case, we derive for the absorption the expression
    \begin{equation}
        Z=\frac{C_{2}(\int\mathcal{R}^{-3}dt-\int\mathcal{R}^{-3}(\exp \{-2\int 3\theta \psi \kappa T_{H}^{-2}\tau dt\})dt)}{C_{1}+C_{2}\int \mathcal{R}^{-3}dt}   \label{eq:deka}
    \end{equation}
As we have already argued, this is always much smaller than 1.

\section{Damping of short waves in Various Viscous Cosmic Fluids}

To derive concrete numbers, our strategy is to compare $\tau$ with $T_{H}$ and then calculate Z defined by (\ref{eq:oktw}). $T_{H}$ was taken as $T_{H}=2t$, 
and $\tau$, $T_{H}$, 
and Z were expressed as functions of the temperature T. We are dealing with radiative fluids, where viscosity arises due to failure of perfect thermal 
equilibrium between matter and highly relativistic particles, like photons and neutrinos. The mean free time was calculated as the inverse of 
the product of the velocity of light with the relevant cross section, $\sigma$, and with the particle density n of the particles of matter which interact 
with the radiation: $\tau =(\sigma n c)^{-1}$. In the 
case of Quark Gluon plasma, the expression of the mean free time 
was taken from Thoma 1991. All the scenarios are realised in the Radiation Dominated Era.
$t_{pl}$ denotes the Planck time, $m_{pl}, m_{p}, m_{e}, m_{\mu}$ the Planck, proton, electron, muon masses respectively, $\alpha$ the fine structure 
constant, k the Boltzmann constant while $L$ is the product of the reduced Hubble parameter with the baryon density parameter: $L= \Omega_{B} h_{r}^{2}$. 
$h_{r}$ is a dimensionless number between 0.4 and 1 that represents the uncertainty to the observed value of $T_{H}^{-1}$ due to systematic errors. $\Omega_{B}$ 
is the ratio of the baryon density of the Universe over the critical density. We have adopted for L the value $2.5\cdot 10^{-2}$.

The time---temperature relation is given by 
     \begin{equation}
          t=0.3(g_{\ast})^{-1/2}(\frac{m_{pl}c^{2}}{kT})^{2}t_{pl}     \label{eq:dwdeka}
     \end{equation}
(see Kolb and Turner 1990) where $g_{\ast}$ denotes the relativistic degrees of freedom (number of effectively massless degrees of freedom, $mc^{2} \ll kT$), 
and varies with time.

\subsection{Quark Gluon Plasma}

According to Mendez et al 1997, the period of interest is $(10^{27} \geq T \geq 10^{24})K$. Taking the expression of the mean free path 
from Thoma 1991, the mean free time  of quarks is given by
     \begin{equation}
         \tau_{q} =3.8\cdot 10\frac{m_{pl}c^{2}}{kT}t_{pl}      \label{eq:evteka}
     \end{equation}
reducing to the expression
     \begin{equation}
         \tau_{q} =1.44\cdot 10^{-20}\frac{10^{10}K}{T}s      \label{eq:evtekatovos}
     \end{equation}
Similarly, for gluons 
     \begin{equation}
         \tau_{g} =1.3\cdot 10\frac{m_{pl}c^{2}}{kT}t_{pl}      \label{eq:embevteka}
     \end{equation}
reducing to the expression
     \begin{equation}
         \tau_{g} =4.83\cdot 10^{-21}\frac{10^{10}K}{T}s      \label{eq:embevtekatovos}
     \end{equation}
The relation between time and temperature reduces to
     \begin{equation}
          t=3.2\cdot 10^{-1}(\frac{10^{10}K}{T})^{2}s     \label{eq:dwdekatovos}
     \end{equation}
$g_{\ast}$ is $106.75$ for the quark gluon plasma era. By considering 3 generations of quarks and 8 kinds of gluons, we found $\kappa =10^{-2}$, $\psi =14$ 
and used $\tau \sim 10^{-20}\frac{10^{10}K}{T}s$. We took 
$\theta = 4/15$ after consulting Thoma 1991. Then, $Z \sim 4.13\cdot 10^{-5}$. Equation (\ref{eq:evadis}) gives a shear viscosity coefficient 
$\xi =7\cdot 10^{-25}K^{-3}T^{3}\, g\, cm^{-1}\, s^{-1}$, while the viscosity coefficient resulting by a straightforward use of the expression of $\xi$ 
found in Thoma 1991 is 
$\xi =4\cdot 10^{-24}K^{-3}T^{3}\, g\, cm^{-1}\, s^{-1}$. There is a good agreement (the ratio of the shear viscosity coefficient of Thoma 1991 to our shear viscosity 
coefficient is at most 5). The absorption due to this mechanism is small, but there is a significant departure of our result 
from the one given in Mendez et al 1997, which is $Z\sim 10^{-7}$. This can be explained from the fact that in Mendez at al 1997 the shear viscosity 
coefficient used is given as $\xi = 1.88\cdot 10^{-26}K^{3}T^{3} g\, cm^{-1}\, s^{-1}$. For the derivation of the latter, Mendez et al 1997 also quote 
Thoma 1991, and they use the same assumptions as we do.

An interesting feature of this absorption mechanism is that the quantity $AT_{H}^{-2}\tau$ (thus, the shear viscosity coefficient) is proportional to 
$T^{3}$, decreasing with time! This is the only considered medium having this behaviour. Large deviations from perfect thermal equilibrium occur in the 
beginning of this era and not its end, unlike the other considered media. Those deviations decrease and the fluid approaches perfect thermal equilibrium 
in the course of time. Viscosity is important towards the beginning of this era, when the density is large and the mean free time is small. As can be seen 
from (\ref{eq:eva}), the density and the mean free time are the parameters to determine viscosity. In the course of time, the decrease of density is 
accompanied by increase of the mean free time. Depending on the fluid, either one or the other primarily determine the viscosity. Since viscosity in the 
quark gluon plasma is greater at large densities, one expects that the collisions that cause the transport of momentum do not involve great momentum 
transfer between particles, but frequently take place and cause the dissipation to occur.

\subsection{Electron Neutrino Mixture}

Two kinds of neutrinos are of main concern for this model, the muon and the electron types, and their antineutrinos. Muon type neutrinos decoupled when the 
muons annihilated at $T \simeq 1.2\cdot 10^{11}K$, since their reaction rate is sensitive to the presence of muons (Weinberg 1972). Electron type neutrinos 
decoupled later, when $T \simeq 10^{10}K$ (time of electron---positron annihilation) because their reaction rate is sensible to the 
presence of electrons. Concerning electron and muon type neutrinos and antineutrinos, electrons, positrons and photons to contribute in the energy density 
of this fluid, $\kappa_{\nu} \equiv \kappa_{\nu_{e}}=\kappa_{\bar{\nu_{e}}}=\kappa_{\nu_{\mu}}=\kappa_{\bar{\nu_{\mu}}} = 10^{-1}$. $\theta =4/15$, in 
agreement to Weinberg 1971.

The period of interest is $(10^{12} \geq T \geq 10^{10})K$. Taking into account the cross section of weak interactions and the electron density 
(de Groot et al 1980), the mean free time of electron type neutrinos and antineutrinos is given by 
     \begin{equation}
         \tau_{\nu_{e}} =2.8\cdot 10^{11}(\frac{m_{pl}}{m_{p}})^{4} (\frac{m_{pl}c^{2}}{kT})^{5}t_{pl}               \label{eq:dekatriaa}
     \end{equation}
reducing to
     \begin{equation}
         \tau_{\nu_{e}} =(1.47\cdot 10 \frac{10^{10}K}{T})^{5}s               \label{eq:dekatriaatovos}
     \end{equation}
For muon type neutrinos and antineutrinos,
     \begin{equation}
         \tau_{\nu_{\mu}} =3.9\cdot 10^{10}(\frac{m_{pl}}{m_{p}})^{4} (\frac{m_{pl}c^{2}}{kT})^{5}\exp \{\frac{m_{\mu}c^{2}}{kT}\} t_{pl}   \label{eq:dekatriab}
     \end{equation}
reducing to
     \begin{equation}
         \tau_{\nu_{\mu}} =2.1\cdot 10^{10}(\frac{10^{10}}{T})^{5}\exp \{\frac{1.23\cdot 10^{12}K}{T}\}   \label{eq:dekatriabtovos}
     \end{equation}
The time---temperature relation is 
     \begin{equation}
          t=1.1(\frac{10^{10}K}{T})^{2}s     \label{eq:dekatesseratovos}
     \end{equation}
The ratio $\tau_{\nu_{\mu}} / \tau_{\nu_{e}}=1.41\cdot \exp \{1.23\cdot 10^{12}K/T\}$ increases rapidly in the period 
$(10^{12} \geq T \geq 1.2\cdot 10^{11})K$ in a range of half up to three orders of magnitude. Therefore, we shall study the behaviour of the fluid in 
two subperiods:
\subsubsection{Subperiod $(10^{12} \geq T \geq 1.2\cdot 10^{11})K$}
Due to the rapid increase of $\tau_{\nu_{\mu}} / \tau_{\nu_{e}}$, it is more convenient to express the shear viscosity coefficient through (\ref{eq:evadis}) 
and compare the average mean free time 
      \begin{equation}
           \bar{\tau} = 2\kappa_{\nu} \tau_{\nu_{e}}(1+ 1.41 \exp \{\frac{m_{\mu}c^{2}}{kT}\})   \label{eq:embolimn}
      \end{equation}
with $T_{H}$. The contribution of muon type neutrinos and antineutrinos to viscosity is greater than that of their electron type counterparts. 
This is due to the faster growth of $\tau_{\nu_{\mu}}$ in this subperiod. The shear viscosity coefficient results to be 
$3.17\cdot 10^{35}K(1+ 1.41 \exp \{1.23\cdot 10^{12}K /T\})T^{-1} g\, cm^{-1}\, s^{-1}$.
\subsubsection{Subperiod $(1.2\cdot 10^{11} \geq T \geq 10^{10})K$}
From now on, only electron neutrinos and antineutrinos contribute to the shear viscosity. We use (\ref{eq:evadis}) with $\psi =2$ and the mean free time 
$\tau_{\nu_{e}}$. The resulting shear viscosity is $3.17\cdot 10^{35}K\, T^{-1} g\, cm^{-1}\, s^{-1}$, in good agreement to the shear viscosity resulting 
from deGroot et al 1980, which is $2.68\cdot 10^{35}K\, T^{-1} g\, cm^{-1}\, s^{-1}$.

Dissipation is $Z \sim 4.11\cdot 10^{-2}$ from both periods. The muon neutrino and antineutrino contribution turns to be of the same order, but slightly 
larger than that of their electron type counterparts. This is due to the faster growth of $\tau_{\nu_{\mu}}$ being compensated by the shorter time period of 
contribution of the muon type neutrinos and antineutrinos to viscosity. We are not aware of anyone having reached a similar or a contradicting result. Neglection of the 
muon neutrino and antineutrino contribution, leads to $Z_{\nu_{e}} \sim 6.19\cdot 10^{-3}$.

Mendez et al 1997 use for the shear viscosity coefficient the expression $\xi = 8.79\cdot 10^{33}K\, T^{-1}\, g\, cm^{-1}\, s^{-1}$  
(they refer to deGroot et al 1980 as the source of their derived shear viscosity 
coefficient as well). This can explain a departure of two orders of magnitude between their result and ours, but cannot justify the actual departure of 
seven orders of magnitude. Indeed, $Z \sim 5\cdot 10^{-9}$ in Mendez et al 1997. These authors have not considered the contribution of the muon neutrinos, 
but this can not explain the difference.

\subsection{Thomson Scattering}

The period of interest is $(10^{9} \geq T \geq 3\cdot 10^{3})K$ (Padmanabham 1993). The mean free time in Mendez et al 1997 is a result of the zero 
chemical potential approximation (also applied to the previous mechanisms). Nevertheless, in these stages of the radiation dominated era, prior to 
recombination, this assumption seems to be not valid, and the hydrogen and baryon abundances should be taken into account. Indeed, setting this mean free 
time equal to the Hubble time, one will reach the result that decoupling occured at some temperature between 
$10^{9}K$ and $10^{8}K$. Following up to a point Kolb and Turner 1990 (the Thomson cross section and number density 
were taken from Weinberg 1972), we adopted the mean free time 
      \begin{equation}
         \tau = \frac{C_{1}S_{1}^{3/2}e^{S_{2}}}{-1+(1+C_{2}S_{3}^{-3/2}e^{S_{2}})^{1/2}}t_{pl}      \label{eq:dekapevte}
      \end{equation}
with $C_{1}=7.56\alpha^{-2} (\frac{m_{e}}{m_{pl}})^{1/2}$, $C_{2}=10^{-7}$, $S_{1}=\frac{m_{pl}c^{2}}{kT}$, $S_{2}=\frac{(m_{p}+m_{e}-m_{H})c^{2}}{kT}$,
$S_{3}=\frac{m_{e}c^{2}}{kT}$.  
(\ref{eq:dekapevte}) reduces to 
      \begin{equation}
         \tau \approx \frac{4.2\cdot 10^{-2}K^{3/2}(T)^{-3/2}\exp (\frac{1.6\cdot 10^{5}K}{T})}{-1+(1+2.2\cdot 10^{-22}K^{-3/2}(T)^{3/2}\exp (\frac{1.6\cdot 10^{5}K}{T}))^{1/2}}s      \label{eq:dekapevtetovos}
      \end{equation}

For $(3.4\cdot 10^{3} > T \geq 3\cdot 10^{3})K$, our mean free time reduces to $\tau =2.8\cdot 10^{9}K^{9/4}T^{-9/4} \exp \{7.9\cdot 10^{4}K\, T^{-1}\}s$, in 
good agreement to $\tau = 8.7\cdot 10^{9}K^{9/4}T^{-9/4} \exp \{8\cdot 10^{4}K\, T^{-1}\}s$ which holds for temperatures close to photon decoupling 
(Padmanabham 1993).

The time---temperature relation is reducing to 
      \begin{equation}
         t \approx 2(\frac{10^{10}K}{T})^{2}s    \label{eq:dekaeksitovos}
      \end{equation}

Absorption is more significant towards the end of the period of consideration ($T<3.6\cdot 10^{3}K$), when departures from perfect thermal equilibrium become 
large. The main contribution is from this subperiod. $\kappa = 7\cdot 10^{-1}$, $\psi =1$ and $\theta = 4/15$.   
We have calculated $Z \sim 7.14\cdot 10^{-4}$. Our result agrees with that of Mendez et al 1997 , who give $Z \sim 2\cdot 10^{-3}$. We cannot explain 
this agreement in results, since we have adopted completely different assumptions. We abandon the zero chemical potential approximation, and in Mendez et 
al 1997 this mechanism is considered only up to $4\cdot 10^{3}K$. Their adoption of zero chemical approximation gives physically wrong results (decoupling 
of photons between $10^{9}K$ and $10^{8}K$, and also a damping due to Thomson scattering six orders of magnitude greater than for Compton scattering, even 
though the latter is a more efficient thermalising mechanism). Besides, their shear viscosity coefficient 
$8\cdot 10^{-19}K^{-5/2}T^{5/2} \exp\{6.02\cdot 10^{9}K\, T^{-1}\}$ 
could never lead to a result such as theirs, because the exponential blows up towards the end of the era of applicability of Thomson scattering, even 
within the temperature range adopted in Mendez et al 1997. 

\subsection{Compton scattering}

The applicability of this mechanism is within $(10^{9} \geq T \geq 5.8\cdot 10^{4})K$. Using the cross section given in Padmanabham, the mean free time turns 
out to be
      \begin{equation}
         \tau =\frac{C_{3}S_{1}^{5/2}e^{S_{2}}}{-1+(1+C_{2}S_{3}^{-3/2}e^{S_{2}})^{1/2}}t_{pl}      \label{eq:dekaepta}
      \end{equation}
where \( C_{3}=7.56\alpha^{-2} (\frac{m_{e}}{m_{pl}})^{3/2} \). This expression is greatly simplified to
      \begin{equation}
          \tau \approx 2\cdot 10^{30}K^{4}T^{-4}s       \label{eq:dekaeptatovos}
      \end{equation}
after a Taylor expansion of the square root in the denominator. This is in good agreement to $\tau = 2.17\cdot 10^{31}K^{4}T^{-4}s$ (Padmanabham 1997). Our 
shear viscosity coefficient is $\xi =5.6\cdot 10^{14}\, g\, cm^{-1}\, s^{-1}$, roughly two orders of magnitude smaller than the 
$\xi=4.37\cdot 10^{16}\, g\, cm^{-1}\, s^{-1}$ derived by using the mean free time of Padmanabham 1993. The result is $Z \sim 6\cdot 10^{-2}$, presenting 
a departure of seven orders of magnitude from the $Z \sim 10^{-9}$ of Mendez et al 1997. This can be explained by their use of 
$\xi=1.72\cdot 10^{5}\, g\, cm^{-1}\, s^{-1}$ (they quote Padmanabham 1993).

The damping due to this mechanism is two orders of magnitude greater than that of Thomson scattering. This is in agreement with the fact that 
Compton scattering is a much more efficient thermalising mechanism than Thomson scattering (Padmanabham 1993).

\section{Damping of Density Perturbations}

The treatment of density perturbations in the non viscous case (for example Grishchuk 1994) shows that they interact with the background gravitational field in the 
same manner as gravitational waves. Since one expects that viscosity will always act against the deformation of the medium produced by a perturbation  
independently of the nature of the perturbation, one expects that density perturbations should have a behaviour similar to gravitational waves in presence 
of viscosity as well. In fact,this should be true for a certain range 
of wavelengths, since we know that density perturbations of smaller wavelengths are washed away. But in the longer  
wavelength limit one could still expect that the behaviour of density perturbations and gravitational waves should be quite similar. Of course, this issue 
requires a more rigorous treatment, following the same steps we have already performed for gravitational waves.

\section{Discussion of results}

The main result of this study has been that gravitational waves which are longer than the Hubble radius today are not practically affected by dissipation, 
and, consequently, there has been no damping in their amplitude.Thus, these modes are good candidates for the production  
of the CMBR large angular scale anisotropy, although a study at the quantum level is required in order to investigate whether viscosity affects more delicate 
properties such as squeezing.

For shorter waves, the picture is different. These modes have been affected by damping more severely than the longer ones. The absorptions we calculated 
for shorter wavelengths are given in the table of results. Compton scattering is the most efficient damping mechanism, Thomson scattering and viscosity due 
to electron neutrino mixture are of comparable efficiency, while the effect of quark gluon plasma in the damping of gravitational waves is much smaller.
In the electron neutrino mixture, it seems that the muon type neutrinos and antineutrinos play a slightly more important role than the electron type ones 
(they both give same order of magnitude damping, with the numerical coefficients slightly in favour of muon type neutrinos and antineutrinos).

Thus, the more time a wave has been within the Hubble radius, the more damped will its amplitude be.The ones longer than today's Hubble radius are not 
affected at all.

There is a large departure of our results from those of Mendez et al 1997, possibly 
due to numerical disagreements. Our estimates of damping are several orders of magnitude higher than the ones of 
Mendez et al. According to these authors, these results are upper limits of damping, and are given in Table 4. In the case of Thomson Scattering, they use 
the zero chemical potential approximation, which as we have shown should be reconsidered, and apply this mechanism for a shorter period of time. In the case 
of other mechanisms, 
the source of our disagreement lies elsewhere. We have used the same references for the 
expressions of mean free paths, cross sections and number densities, and the same assumptions. Their approach 
is much more complicated than ours: they set the Einstein equations in quasi-Maxwellian form and retain in their formulas the shear viscosity coefficient. 
Then, they solve the resulting differential equations either analytically or numerically. If one calculates the shear viscosity coefficients directly from 
the references, these turn out to be several orders of magnitude larger than those used in Mendez et al 1997. These departures can explain the 
differences of our results with the results of Mendez et al 1997.

\section{Acknowledgements}

The author is supported by the Foundation of State Scholarships of Greece.\\
The author expresses his gratitude to Professor LP Grishchuk for his valuable advice and enlightening conversations.\\
The help of Assistant Professor T Christodoulakis was great in the numerical integration of the resulting integral for the case of Thomson Scattering.\\
The author wishes to thank the members of the Relativity Group of the University 
of Wales Cardiff for interesting discussions and remarks.

\begin{table}
   \begin{tabular}{|c|c|c|}    \hline
   Viscous Cosmic Medium     &Era of Application                         &Damping \\  \hline
   Quark Gluon Plasma        &$(10^{27} \geq T \geq 10^{24})K$           &$Z \sim 1.55\cdot 10^{-4}$ \\  \hline 
   Electron Neutrino Mixture &$(10^{12} \geq T \geq 10^{10})K$           &$Z \sim 4.11\cdot 10^{-2}$ \\  \hline
   Thomson Scattering        &$(10^{9} \geq T \geq 3.1\cdot 10^{3})K$    &$Z \sim 7.14\cdot 10^{-4}$ \\  \hline
   Compton Scattering        &$(10^{9} \geq T \geq 5.8\cdot 10^{4})K$    &$Z \sim 6.14\cdot 10^{-2}$ \\  \hline
   \end{tabular}
   \caption{TABLE OF RESULTS}
\end{table}

\begin{table}
   \begin{tabular}{|c|c|c|}    \hline
   Viscous Cosmic Medium     &Era of Application                       &Damping \\  \hline
   Quark Gluon Plasma        &$(10^{27} \geq T \geq 10^{24})K$         &$\sim 10^{-7}$        \\  \hline 
   Electron Neutrino Mixture &$(10^{12} \geq T \geq 10^{10})K$         &$\sim 5\cdot 10^{-9}$ \\  \hline
   Thomson Scattering        &$(10^{9} \geq T \geq 4\cdot 10^{3})K$    &$\sim 2\cdot 10^{-3}$ \\  \hline
   Compton Scattering        &$(10^{9} \geq T \geq 5.8\cdot 10^{4})K$  &$\sim 10^{-9}$        \\  \hline
   \end{tabular}
   \caption{TABLE OF RESULTS OF MENDEZ ET AL 1997}
\end{table}

\section{References}
de Groot SR et al 1980 "Relativistic Kinetic Theory" (Amsterdam:North Holland)\\
Grishchuk LP 1993 Class Quantum Grav 10 2449\\
Grishchuk LP 1994 Phys Review D 50 7154\\
Grishchuk LP and Polnarev AG 1980 "General Relativity And Gravitation" vol 2 ed A Held (New York: Plenum)\\
Hawking SW 1966 Astrophys J 145 544\\
Kolb EW and Turner MS 1990 "The Early Universe" (New York: Addison---Wesley)\\
Landau LD and Lifshitz EM 1966 "Fluid Mechanics" (Pergamon Press)\\
Lifshitz EM and Khalatnikov IM 1963 Adv Phys 12 185\\
Mendez V et al 1997 Class Quantum Grav 14 77\\
Padmanabham T 1993 "Structure Formation In The Universe" (Cambridge: Cambridge University Press)\\
Reif F 1965 "Fundamentals of Statistical and Thermal Physics" (New York: McGraw---Hill)\\
Tabor D 1970 "Gases, Liquids And Solids" (Glasgow:Bell and Bain)\\
Thoma MH 1991 Phys Lett B 269 144\\
Weinberg SW 1971 Astrophys J 168 175\\
Weinberg SW 1972 "Gravitation And Cosmology" (New York:Wiley)\\

\end{document}